\documentclass[twocolumn,preprintnumbers,prl,showkeys,showpacs]{revtex4}
\usepackage{amsmath}
\usepackage{amssymb,amsfonts}

\makeatletter
\renewcommand{\section}{\@startsection{section}{1}{0mm}{-2\baselineskip}
{-0.5\baselineskip}{\normalfont\normalsize\bf}}
\renewcommand{\subsection}{\@startsection{subsection}{2}{0mm}{-\baselineskip}
{-0.5\baselineskip}{\normalfont\normalsize\bf}}
\newcommand{\d@t}{.}
\makeatother

\newcommand{\p}{\partial}

\def\bea{\begin{eqnarray}}
\def\eea{\end{eqnarray}}
\def\be{\begin{equation}}
\def\ee{\end{equation}}
\def\a{\alpha}
\def\b{\beta}
\def\g{\gamma}

\def\r{\rho}
\def\om{\omega}
\newcommand{\Zint}{\mathbb{Z}}
\newcommand{\Real}{\mathbb{R}}

\begin{document}

\title{Quantum Cosmology and Conformal Invariance}


\author{B. Pioline}
\email{pioline@lpthe.jussieu.fr}
\affiliation{LPTHE, Universit\'es Paris VI et VII, 4 pl Jussieu, \\
75252 Paris cedex 05, France}

\author{A. Waldron}
\email{wally@math.ucdavis.edu}
\affiliation{Department of Mathematics,\\
University of California at Davis, CA 95616, USA}

\preprint{LPTHE-02-42, {\tt hep-th/0209044}}  

\begin{abstract}
According to Belinsky, Khalatnikov and Lifshitz, gravity
near a space-like singularity reduces to a set of decoupled one-dimensional 
mechanical models at each point in space. We point out that these models
fall into a class of conformal mechanical models first introduced 
by de Alfaro, Fubini and Furlan (DFF). The deformation used by DFF 
to render the spectrum discrete corresponds to 
a negative cosmological constant. The wave function of the Universe 
is the zero-energy eigenmode of the Hamiltonian, or 
the spherical vector of the representation of the
conformal group $SO(1,2)$. A new class of conformal quantum mechanical 
models with enhanced ADE symmetry is constructed, 
based on the quantization of nilpotent coadjoint
orbits.
\end{abstract}



\maketitle
While cosmological singularities have so far eluded any satisfactory treatment
in quantum gravity, insight into their generic properties is
afforded by the  classical 
analysis of Belinsky, Khalatnikov and Lifshitz (BKL)~\cite{bkl}.
Under a self-consistent hypothesis of decoupling of the
dynamics at nearby points upon approaching a space-like singularity, they
show that a generic solution of four-dimensional Einstein gravity
exhibits a chaotic oscillatory behavior near the big bang (or big
crunch). This subject has attracted much recent attention,
with the discovery that the chaotic properties of the dynamics are
tied to the hyperbolicity of the Kac-Moody algebra underlying the 
billiard geometry on which the motion takes 
place~\cite{Damour:2000hv,Damour:2001sa}.
This is in particular the case of all models descending from
eleven-dimensional supergravity or $N=1$ supergravity in ten 
dimensions~\cite{Damour:2000wm}.
The spatial gradients, although subleading at the singularity,
are also described by the Kac-Moody structure 
\cite{Damour:2002cu}.

Our aim in this letter is twofold. Firstly, we point out an important
feature of the one-dimensional 
model for the gravity modes at each decoupled point: 
it exhibits the conformal invariance
found in one-dimensional quantum mechanics by de Alfaro, Fubini 
and Furlan (DFF)~\cite{dff}. 
Conformal quantum mechanical models 
have already appeared in black hole physics~\cite{claus,volo},
yet their relevance to cosmology seems to have been hitherto unnoticed.
More specifically, the Wheeler-DeWitt equation 
is but the Schr\"odinger equation considered by DFF, 
restricted to zero energy. The wave function of the universe thus lies
at the bottom of a continuum of delta-function normalizable states.
Retaining the effect of a negative cosmological constant
discretizes the spectrum while preserving conformal invariance. 
In mathematical terms, the wave function of the universe is therefore 
the spherical 
vector of the representation of the conformal group. Secondly, we construct
a new class of conformal quantum mechanical 
models, where the conformal group is enhanced to an ADE 
non-compact group. In these models, the spherical vector,
and hence the wave function of the universe is known exactly.

We start with $n$+1 dimensional Einstein gravity with
a cosmological constant and reduce down to
a 0+1 dimensional system.
Parameterizing the metric as
\be
ds^2=-\left[\frac{\eta(t)}{V(t)}\right]^2 dt^2 + V^{2/n}(t)\,  
\hat g_{ij}(t)\,  dx^i dx^j\, ,
\ee
the Einstein-Hilbert action reduces to
\be
\begin{split}
\label{particle}
\int &dt~d^nx~\sqrt{-g}\, (R-2\Lambda)\, =\,  \\
&\int ~dt \left\{ \frac{1}{2\eta} \left[
-\frac{2 (n-1)}{n}\dot V^2 
+V^2\, \dot U^M G_{MN} \dot U^N \right]
-2\Lambda \eta \right\}\, 
\end{split}
\ee
where $V$ denotes the volume of the spatial metric
and $U^M$ coordinatize the symmetric space $S=Sl(n)/SO(n)$,
with constant negative curvature, and 
homogeneous metric $d U^M G_{MN} d U^N:=-\frac{1}{2}
d{\hat{g}}_{ij} d{\hat g}^{ij}$ (with $\det \hat g=1$). In this form,
we recognize the action for a ``fictitious''
point particle, with mass squared $m^2=4\Lambda$, tachyonic 
for negative (AdS) $\Lambda$, propagating on 
a Lorentzian cone with base $S$ and metric
\be
\label{conen}
d\sigma^2 = - \frac{ 2 (n-1)}{n}\, dV^2 + V^2 dU^M G_{MN} dU^N\, .
\ee
The rescaled lapse $\eta$ plays the role of an einbein gauge field
enforcing invariance under general time reparameterizations.
The appearance of the volume $V$ with a negative kinetic term
suggests its use as a ``cosmological time''.
Indeed, it develops with
time as $[(V/\eta){d}/{dt}]^2 V = [2 n \Lambda/(n-1)] V$,
while the particle follows geodesics on the cone and hence
on its base $S$. A similar reduction in the presence of
extra scalar and gauge fields would yield a cone over an enlarged
homogeneous space (e.g. $SO(n,n)/[SO(n)\times SO(n)]$ in 
the presence of a Kalb-Ramond two-form and dilaton).

Since the conical moduli space~\eqref{conen} admits an homothetic 
Killing vector
$V\p_V$,  the free particle should exhibit conformal
invariance~\cite{volo}. This is easily shown by
introducing conjugate momenta $p$ and $P_M$ 
for the canonically normalized volume coordinate
$\rho=\sqrt{8(n-1)V/n}$ and the shape moduli $U^M$.
The Hamiltonian following from the action \eqref{particle} reads
\be
\label{hamconf}
H =  \frac{\eta}{V} \left[ \frac12 p^2 + \frac{4(n-1)}{n \rho^2} \Delta
- \frac{n\Lambda}{4(n-1)} \rho^2 \right] .
\ee
The equation of motion of $\eta$ forces $H$ to vanish,
after which the gauge $\eta=V$ can be imposed.
Here $\Delta=-P_M G^{MN} P_N$ is the quadratic Casimir
of the action of $Sl(n)$ on the homogeneous space
$Sl(n)/SO(n)$ and
has vanishing Poisson bracket withs $\rho$ and $p$. Therefore it 
effectively plays the r\^ole of a 
coupling constant $g=8(n-1)\Delta/n$ for the $1/\rho^2$
potential. Indeed at $\Lambda=0$, we recognize in~\eqref{hamconf} 
the Hamiltonian of the conformal mechanical system 
introduced by de Alfaro, Fubini and Furlan~\cite{dff}.
The generators 
\be
\label{so21}
E_+=\frac12\r^2\ ,
\quad D_0=\frac12 \r p\ ,\quad E_-=\frac12 \left( p^2 + \frac{g}{\r^2}
 \right)
\ee
represent the conformal group $SO(1,2)$ in 0+1 dimensions, 
$\{E_+,E_-\}=2D_0,\  \{D_0,E_\pm\}=\pm E_{\pm}$.
For vanishing $\Lambda$, the Hamiltonian $H=E_-$ has 
unit dimension w.r.t. the generator of conformal rescalings $D_0$,
hence the system is conformally invariant. Remarkably,
the introduction of a cosmological term preserves the
action of the conformal group $SO(2,1)$, as it simply
amounts to choosing a different generator $H=E_- - n\Lambda/(2n-2) E_+$ 
as the Hamiltonian. This deformation was considered in
\cite{dff} although without a clear physical motivation. 


Let us now discuss some aspects of the quantum dynamics in this 
toy model of a cosmological singularity. While we understand little about 
quantum gravity at a space-like singularity,
we may assume that the decoupling of nearby points still holds and
work in a minisuperspace truncation. Replacing canonical momenta
by their Schr\"odinger representation $p\to i\p/\p\rho\ , P_M \to
i \p/\p{U^M}$, the Hamiltonian constraint 
\eqref{hamconf} becomes the Wheeler-DeWitt equation \cite{dw},
\be
\label{wdw}
H \psi = \left( -\frac12 \p^2 + \frac{4(n-1)}{n~\rho^2} \Delta
- \frac{n\Lambda}{4(n-1)} \rho^2 \right) \psi =0\, ,
\ee
acting on wave functions $\psi(\rho,U^M)$, where
$\Delta$ is the quadratic Casimir of the $Sl(n)$ action
on the homogeneous space $S$. As usual, a vanishing
Hamiltonian implies that the wave function $\psi$ is independent
of the time $t$, nevertheless correlations between the volume $\rho$ and
the other observables $U_M$ may be used to set up measurements \cite{dw}. 
One may now recognize the Wheeler-DeWitt equation~\eqref{wdw}
as the Schr\"odinger equation of DFF's conformal quantum mechanics, 
with coupling $g=8(n-1)\Delta/n$,
restricted to zero-energy states. Invariance under the conformal
group $SO(1,2)$ is retained at the quantum level  after resolving
the ordering ambiguity $D_0 \to (\r p+p \r)/4$. (Indeed, 
the requirement of conformal 
symmetry and its extension below, uniquely fix all ordering ambiguities.) 
In particular, the effect of a cosmological constant $\Lambda<0$ is to 
replace the parabolic generator $E_-$ with continuous spectrum
$\Real^+$, by a compact, discrete spectrum, generator $H=E_--
 n\Lambda/(2n-2) E_+$ (for $\Lambda>0$, $H$ has a continuous spectrum). 
It is intriguing that this seemingly favorable case 
corresponds to a {\it tachyonic} fictitious particle.

Despite the formal identity between the WDW and DFF Hamiltonians,
it is worth noting several crucial differences.
Firstly, the WDW equation picks out zero energy modes of $H$ only, so that
the requirement of boundedness from below is no longer necessary.
Indeed, the coupling $g$ in our problem appears to be negative on square 
integrable wave functions on $S$ (for which $\Delta<0$), and so
does the mass term for $\Lambda<0$. This is in fact a standard feature
of canonical gravity, where the conformal factor always appears with
a kinetic term of the ``wrong'' sign \cite{dw}. Similarly, 
in a traditional quantum mechanics set-up, one usually requires
states to have a finite $L_2$ norm around $\rho=0$
\cite{dff}. When $\rho$ is viewed as a cosmological time, the requirement
of square normalisability is no longer sensible (it can however be useful
to select recollapsing universes \cite{dw}).  The analogy of
\eqref{wdw} with a massive Klein--Gordon equation would suggest instead to
consider the Klein-Gordon norm on space-like slices of fixed $\rho$
(``third'' quantization may be used to cure the non-positive
definiteness~\cite{third}, although its interpretation remains unclear.)
At any rate, our current understanding of cosmological singularities
does not allow us to specify the boundary conditions reliably,
we therefore proceed without further ado.

A few comments are in order about the chaotic properties
of these cosmological models. Firstly, our discussion was carried out 
for free Kasner flights, in the absence of the potential terms
coming from spatial gradients. As one approaches the
space-like singularity, the potential terms behave as infinitely steep
reflection walls \cite{Misner}. 
It is possible that their effect could be mimicked by modding out
by a discrete symmetry group (e.g, the walls exchanging the various
radii are included in a $Sl(n,\Zint)$ subgroup of $Sl(n)$ acting
on $S$). This group is however of a rather wild nature, as it should
contain the Weyl group of an hyperbolic Kac-Moody algebra~\cite{Damour:2000hv,
Damour:2001sa} and it is unclear at this stage how to describe the 
geodesic motion on such an object.  One may speculate
that the universal $SO(2,1)$ subalgebra 
uncovered in~\cite{Nicolai:2001ir} in general hyperbolic Kac-Moody algebra
may play a role in implementing conformal invariance.

We now present a construction of a class of conformally
invariant quantum systems based on the quantization of nilpotent
coadjoint orbits of finite Lie groups. In contrast to generic orbits,
nilpotent ones are especially interesting as they possess fewer or no
free parameters. Such conformal systems were first 
found in the course of constructing of theta series 
for non-symplectic groups
\cite{kpw}, and a particular example was given independently 
in~\cite{gkn} for the minimal representation of
$E_8$. For simplicity, we shall illustrate it on the simplest non-trivial
case, $D_4$, which will coincide with the model~\eqref{wdw}
for $2+1$ gravity.

The classical phase space of our systems  arises from
the coadjoint orbit of a nilpotent element of smallest order 
in a finite simple Lie algebra $G$. This element can be conjugated into
the generator associated to the lowest root $E_{-\omega}$. The
$Sl(2)$ subalgebra generated by  $\{ E_{\omega}, D_{\omega}, E_{-\om} \}$
with $D_\omega=[E_{\om},E_{-\om}]$
will be the conformal group in 0+1 dimensions, and $E_{-\omega}$ be chosen
as the Hamiltonian. The Cartan generator $D_{\omega}$ grades the
algebra into 5 subspaces, $G=G_{-2}\oplus G_{-1}\oplus G_{0}\oplus G_{1}
\oplus G_{2}$, such that the top and bottom subspaces are one-dimensional,
$G_{\pm 2}=\Real E_{\pm \omega}$. The coadjoint orbit of $E_{-\omega}$ can
be parameterized by $P\backslash G=\Real H_\omega \oplus G_1 \oplus G_2$, 
where $P$ is the stabilizer of $E_{-\omega}$ under the coadjoint
action. The level-one space
$G_1$ is a Heisenberg algebra, which can be diagonalized in
the form $[E_{\b_i},E_{\g_j}]=\delta_{ij} E_{\om}$. We can thus represent
these generators as canonical coordinates and conjugate momenta,
$E_{\b_i}=y p_i,\ E_{\g_i}= x_i\ ,E_{\om}=y$.
The maximal subalgebra $H$ in $G$ commuting with $SO(2,1)$  
lies in $G_0$ and acts linearly
as canonical transformations on the coordinates and momenta $\{x_i,p_i\}$,
leaving $y$ invariant. The choice of a polarization of $G_1$ 
into coordinates and momenta further breaks $H$ to a subgroup $H_l$
acting linearly on the coordinates. A standard polarization is to choose
for $\b_0$ the simple root
to which the affine root $\omega$ attaches on the Dynkin diagram of $G$
and for $\beta_{i>0}$ the positive roots such that $\langle \beta_0,\beta_i
\rangle=1$~\cite{kazhdan}. 
The canonical momentum associated to $y$ is obtained  in turn 
from the Cartan generator associated to $\beta_0$, 
$D_{\beta_0}:=[E_{\b_0},E_{-\b_0}]=py-p_0 x_0$. So as to bring
the generator $E_{\om}$ to the conformal quantum mechanics form
\eqref{so21}, we make a canonical transformation 
\be
\label{cano}
y=\frac{\rho^2}{2},\ 
x_i=\frac{\rho q_i}{2},\ 
p=\frac{1}{\r}p_\r- \frac{1}{\r^2}q_i \pi_i,\ 
p_i=2\frac{\pi_i}{\r}
\ee
The generators of the $SO(2,1)$ subalgebra take the form
\be
\label{so212}
{E_{\om}}=\frac12\rho^2\ ,\quad
D_{\om}=\rho p_\r\ ,
\quad E_{-\om}=\frac12 \left( p^2 + 
\frac{4\Delta}{\rho^2} \right)
\ee
where $\Delta$ is, up to an additive constant, the quadratic Casimir of
$H$, hence a quartic invariant of the coordinates and momenta $\{q_i,\pi_i\}$. 
Finally, the symmetry is enhanced from $SO(2,1)\times H$ to all of $G$
using two discrete generators: (i) the Fourier transform on all positions
$q_i$ at once, corresponding to the longest word in the Weyl group,
and (ii) the Weyl reflection with respect to $\beta_0$, 
acting on wave functions as
$W\psi(\rho, q_0, q_i) = e^{-\frac{I_3(q_i)}{2q_0}}
\psi\left( \sqrt{-\rho q_0}, \sqrt{-\rho^3/ q_0}, 
\sqrt{-\rho q_i/q_0} \right)
$
where $I_3$ is the cubic invariant of the
positions under the linearly realized $H_l$.
The compatibility between the two actions, and indeed the whole
construction, relies heavily on the invariance of 
the non-Gaussian character $\exp(i I_3(x_i)/x_0)$ under Fourier 
transform \cite{kazhdan}.

As an example, the algebra $D_4$ decomposes as $1_{-2}\oplus (2,2,2)_{-1}
\oplus [(1,1,1)+(3,1,1)+{\rm perm}]_0\oplus (2,2,2)_1\oplus 1_2$ under 
$Sl(2)^3 \times \Real$, where $\Real$ denotes the Cartan generator
$D_{\om}$ associated to the highest root. The coordinates and momenta
correspond to the grade-one space and transform as a $(2,2,2)$
of $H=Sl(2)^3$. They satisfy the Heisenberg algebra
$
[q^{aA\alpha},q^{bB\beta}]=\epsilon^{ab}\epsilon^{AB}\epsilon^{\a\b}
$.
The actions of each $Sl(2)$ factor in $H$ are represented by the 
angular momentum-like operators
$
\Sigma^{\mu}=\sigma^\mu_{\a\b}~~ \epsilon_{ab}~\epsilon_{AB}~
q^{aA\alpha}~q^{bB\beta}
$,
with similar definitions for $s^m$ and $S^M$.
It is easy to check that these generators satisfy the $Sl(2)$ algebra,
$[\Sigma^{\mu},\Sigma^{\nu}]=\epsilon^{\mu\nu\rho} \Sigma^{\rho}$.
The quadratic Casimirs of all three $Sl(2)$'s are identical
and equal to the unique quartic invariant of the $(2,2,2)$ 
representation. The generators of the $SO(2,1)$ subalgebra then read
as in~\eqref{so212} with  $\Delta$ the common quadratic Casimir of the
three $Sl(2)$. To clarify the meaning of the Hamiltonian
in~\eqref{so212}, let us choose a polarization such that the 
$Q^{A\alpha}=q^{1A\alpha}$ are coordinates and $q^{2A\alpha}$ are momenta.
The bispinor $Q^{A\a}$ can be thought of as a vector $Q^I$ of $SO(2,2)$,
parameterized by  three ``polar angles'' $\Omega\in
H_3=SO(2,2)/SO(2,1)=SO(2,1)$ and its length squared 
$\kappa^2=Q^I \eta_{IJ} Q^J$,
where $\eta_{IJ}$ is the signature $(2,2)$ metric.
The quadratic Casimir then 
corresponds to the angular momentum squared on the pseudo-sphere $H_3$, 
{\it i.e.} the Laplacian on $Sl(2)$. Out of the four coordinates $Q^{I}$,
we see that only the three hyperbolic polar angles receive
kinetic terms in the Hamiltonian $E_{-\omega}$, while the radius
$\kappa$ decouples. 

\begin{table*}[t]
\caption{Non-linearly realized symmetry $G$, canonically realized $H$,
representation of positions and momenta under $H$, linearly realized
$H_l$, homogeneous space $S$ including decoupled factors $\Real_\kappa$,
and functional dimension of the Hilbert space for models with enhanced
conformal symmetry.}
\begin{ruledtabular}
\begin{tabular}{llllll}
 $G$ & $H$ & $G_1$ & $H_l$ & $S$ & dim\\
 $A_{n-1}$ & $Gl(n-2)$ & $[n-2]+[n-2]$ & $Sl(n-2)$ &
    $\Real_\kappa\times Sl(n-2)/[Sl(n-3)\times \Real^{n-3}]$ & $n-1$\\ 
 $D_n$ & $Sl(2) \times D_{n-2}$ & $(2,2n-4)$ & $D_{n-2}$  &
    $\Real_\kappa \times {SO(n-2,n-2)}/{SO(n-3,n-2)}$ & $2n-3$\\
 $E_6$ & $Sl(6)$ & $20=[0,0,1,0,0]$ &$SO(3,3)$  & 
    $\Real_\kappa \times SO(3,3)/[SO(3)\times SO(3)]$ & 11\\
 $E_7$ & $SO(6,6)$ & $32$  & $Sl(6)$ 
 &$\Real_\kappa \times Gl(6)/Sp(6)$
& 17\\
 $E_8$ & $E_7$     & $56$  & 
$Sl(8)$  
 & 
$\Real_\kappa \times Sl(8)/Sp(8)$ & 29
\end{tabular}
\end{ruledtabular}
\end{table*}

A natural object in the theory of minimal representations
is the spherical vector, {\it i.e.} the wave function annihilated by all compact
generators $E_{\alpha}+E_{-\alpha}$. For irreducible and so-called
spherical representations, this vector is unique. 
{}From the expression~\eqref{so212}, we see that a
spherical vector corresponds to a zero-energy state of the Hamiltonian\
with a negative ``cosmological constant'', 
{\it i.e.} $H=E_{-\omega}+E_{\omega}$.
For $D_4$, the spherical vector should also be invariant under
the compact $U(1)$ inside each $Sl(2)$ factor, so the coordinates
of the $D_4$ conformal quantum mechanics are effectively valued in
$H_3/U(1)=Sl(2)/U(1)$. 
This is indeed the conformal model~\eqref{wdw} 
of $2+1$ gravity with $S=SO(2)$! Therefore,
supplementing the fields 
$(\rho; U_1,U_2) \in \Real \times Sl(2)/U(1)$ with an angular variable 
and a decoupled radius $\kappa$, the conformal symmetry of reduced 
$D=2+1$ gravity is enhanced
to an $SO(4,4)$ non-compact spectrum generating symmetry. The spherical
vector of the $D_4$ minimal representation has been obtained in 
\cite{kpw}, in the  polarization used here it reads
\be
\psi_{D_4}=\frac{\rho^{3/2}e^{-S}}{S}\ ,\ 
S=\frac12 \sqrt{\rho^4+\rho^2{\rm tr}(Q^t Q)+\kappa^4}
\ee
It would be very interesting if the new coordinate $\kappa^2=
\det(Q^{A\alpha})$, appearing here as a degeneracy label, 
had a cosmological interpretation.


A similar conformal quantum mechanical system can be constructed 
for any finite simple Lie algebra $G$, except for the $SO(2n+1)$ series. 
The details of 
the construction of the minimal representation have been spelled out 
for the simply laced case in~\cite{kpw}, where
the spherical vector has also been obtained.
In order to translate the results of~\cite{kpw} into 
the presentation~\eqref{so212} suitable for interpreting $E_{-\om}$
as the Hamiltonian of a quantum mechanical system, one only needs
to perform the canonical transformation~\eqref{cano}. 
We thus obtain a class of quantum mechanical systems 
with an enhanced conformal symmetry $G$ corresponding to
any finite simple Lie group 
in (non-compact) split real form. 
The field content and symmetries of these models are summarized in Table 1.
A similar construction can also be carried out for other (non-nilpotent)
orbits, or for the non-simply laced groups
$G_2$ and $F_4$, whose minimal representation 
does not possess any spherical vector, 
so that part of the compact symmetries are spontaneously broken. 
An open problem is to identify gravitational theories reducing
to those in the process of dimensional reduction.


In conclusion, conformal quantum
mechanics is relevant 
to the dynamics of gravity at a space-like singularity.
A negative cosmological constant renders the spectrum of 
the Wheeler-DeWitt operator discrete. The
wave function of the universe is then  obtained as the spherical
vector of the representation of the enhanced conformal group. We have
constructed a family of conformal quantum mechanical models with
symmetry enhanced to an arbitrary simple non-compact group $G$ in split
real form. For $G=D_4$, this reproduces the conformal mechanics arising
from the reduction of $2+1$-dimensional gravity, together with a decoupled 
field $\kappa$ yet to be understood. 
An important
question is whether this construction can be extended to infinite
Kac-Moody groups such as $E_{10}$ or $BE_{10}$, which should control
the dynamics of M-theory or the heterotic string at a space-like singularity.
It would be exciting if the dynamics of gravity at a spacelike
singularity was related to the geodesic motion on a coadjoint orbit 
of $E_{10}$, in analogy with conventional incompressible fluid 
hydrodynamics. 

\begin{acknowledgments}
{\it Acknowledgments.} 
It is a pleasure to thank E. Rabinovici, M. Berkooz, H. Nicolai for 
valuable discussions, and the Albert Einstein
Institute, Golm and Max Planck
Institut f\"ur Mathematik, Bonn for hospitality.
Research supported
in part by NSF grant PHY-0140365.
\end{acknowledgments}

\end{document}